\documentclass[10pt]{article}

\usepackage{longtable, booktabs, makecell, multirow, array, xcolor, graphicx}
\usepackage[margin=2cm]{geometry}
\usepackage{caption}

\usepackage{fullpage}
\usepackage{setspace}
\usepackage{parskip}
\usepackage{titlesec}
\usepackage[section]{placeins}
\usepackage{xcolor}
\usepackage{breakcites}
\usepackage{lineno}
\usepackage{hyphenat}
\usepackage[numbers]{natbib}
\PassOptionsToPackage{hyphens}{url}
\usepackage[colorlinks = true,
            linkcolor = blue,
            urlcolor  = blue,
            citecolor = blue,
            anchorcolor = blue]{hyperref}
\usepackage{etoolbox}


\renewenvironment{abstract}
  {{\bfseries\noindent{\abstractname}\par\nobreak}\normalsize}
  {\bigskip}

\titlespacing{\section}{0pt}{*3}{*1}
\titlespacing{\subsection}{0pt}{*2}{*0.5}
\titlespacing{\paragraph}{0pt}{*1.5}{0pt}

\usepackage{authblk}

\usepackage{graphicx}
\usepackage[space]{grffile}
\usepackage{latexsym}
\usepackage{textcomp}
\usepackage{longtable}
\usepackage{tabulary}
\usepackage{booktabs,array,multirow}
\usepackage{amsfonts,amsmath,amssymb}
\providecommand\citet{\cite}
\providecommand\citep{\cite}

\newif\iflatexml\latexmlfalse

\AtBeginDocument{\DeclareGraphicsExtensions{.pdf,.PDF,.eps,.EPS,.png,.PNG,.tif,.TIF,.jpg,.JPG,.jpeg,.JPEG}}

\usepackage[utf8]{inputenc}
\usepackage[greek,english]{babel}

\begin{document}

% \title{Bridging Lipid-deficient Molecular Subtypes from Histopathology to Preoperative CT in Clear Cell Renal Cell Carcinoma}

\title{Multiscale Cross-Modal Mapping of Molecular, Pathologic, and Radiologic Phenotypes in Lipid-Deficient Clear Cell Renal Cell Carcinoma}

% \title{Cross-Modal Mapping of Molecular, Pathologic, and Radiologic Phenotypes in Lipid-Deficient Clear Cell Renal Cell Carcinoma}
% \title{Multiscale Cross-Modal Mapping of Molecular, Pathologic, and Radiologic Signatures in Lipid-Deficient ccRCC}

% \title{Multiscale Cross-Modal Translation of Histopathology into CT Phenotypes in Lipid-Deficient Clear Cell Renal Cell Carcinoma}

% \title{A cross-modal framework translates histopathologic signatures of a lipid-deficient subtype into preoperative CT biomarkers for clear cell renal cell carcinoma}

% \title{A cross-modal framework maps lipid-deficient (DCCD) histopathology to preoperative CT in clear cell renal cell carcinoma}

\author[1]{Ying Cui$^*$}
\author[2]{Dongzhe Zheng$^*$}
\author[3]{Ke Yu$^*$}
\author[4]{Xiyin Zheng}
\author[5]{Xiaorui Wang}
\author[6]{Xinxiang Li}
\author[7]{Yan Gu}
\author[1]{Lin Fu}
\author[1]{Xinyi Chen}
\author[8]{Wenjie Mei$^\dagger$}
\author[1]{Xin-Gui Peng$^\dagger$}

\affil[1]{Nurturing Center of Jiangsu Province for State Laboratory of AI Imaging \& Interventional Radiology, Department of Radiology, Zhongda Hospital, School of Medicine, Southeast University, Nanjing, 210009, China} 
\affil[2]{Department of Mechanical and Aerospace Engineering, Princeton University, Princeton, NJ 08544, USA}
\affil[3]{School of Automation, Southeast University, Nanjing, 210096, China}
\affil[4]{Department of Biomedical Engineering, Columbia University, New York, NY 10027, USA}
\affil[5]{Automation and Electrical Engineering College, Lanzhou University of Technology, Lanzhou, 730050, China}
\affil[6]{The First Affiliated Hospital of USTC, Division of Life Sciences and Medicine, University of Science and Technology of China, Hefei, 230031, China}
\affil[7]{Lianyungang First People's Hospital, 182 Tongguan North Road, Lianyungang, 222002, China}
\affil[8]{School of Robotics and Automation, Suzhou Campus, Nanjing University, 1520 Taihu Road, Suzhou, 215163, China}

%\author[1]{Anonymous Authors}

\begingroup
\let\center\flushleft
\let\endcenter\endflushleft
\maketitle
\endgroup

\vspace{-1em}

{$^*$Contributed equally as the joint first authors. $^\dagger$Contributed equally as the joint last authors.}

{\textbf{Corresponding Authors}: 
Xin-Gui Peng, MD, PhD (Nurturing Center of Jiangsu Province for State Laboratory of AI Imaging \& Interventional Radiology, Department of Radiology, Zhongda Hospital, School of Medicine, Southeast University, Nanjing, 210009, China. Email: \href{mailto:xinguipeng@seu.edu.cn}{xinguipeng@seu.edu.cn}).  Wenjie Mei, PhD (School of Robotics and Automation, Suzhou Campus, Nanjing University, 1520 Taihu Road, Suzhou, 215163, China. Email: \href{mailto:mei.wenjie@nju.edu.cn}{mei.wenjie@nju.edu.cn}). } 

{\textbf{Author Contributions}:  
XGP and WM conceptualized and designed the study and conducted the investigation. YC, DZ, KY, XW, XL, YG, LF, and XC collected the data. YC and DZ analyzed the data. YC, DZ, and KY wrote the original draft. XGP and WM have accessed and verified the data. XGP and WM provided supervision. All authors reviewed and approved the final version of the manuscript. All authors have full access to all the data in the study and have final responsibility for the decision to submit for publication.}

{ \textbf{Conflict of Interest Disclosures}:
The authors have no conflicts to report.} 

{ \textbf{Founding}: This study has received funding from the National Natural Science Foundation of China (82272064, 62403125), the Natural Science Foundation of Jiangsu Province (BK20221461, BK20241283), Zhongda Hospital Affiliated to Southeast University, Jiangsu Province High-Level Hospital Pairing Assistance Construction (zdlyg08).}

\selectlanguage{english}

%\newpage 

\begin{abstract}
Clear cell renal cell carcinoma (ccRCC) exhibits extensive intratumoral heterogeneity on multiple biological scales, contributing to variable clinical outcomes and limiting the effectiveness of conventional TNM staging, which highlights the urgent need for multiscale integrative analytic frameworks. The lipid-deficient de-clear cell differentiated (DCCD) ccRCC subtype, defined by multi-omics analyses, is associated with adverse outcomes even in early-stage disease. Here, we establish a hierarchical cross-scale framework for the preoperative identification of DCCD-ccRCC. At the highest layer, cross-modal mapping transferred molecular signatures to histological and CT phenotypes, establishing a molecular-to-pathology-to-radiology supervisory bridge. Within this framework, each modality-specific model is designed to mirror the inherent hierarchical structure of tumor biology. PathoDCCD captured multi-scale microscopic features, from cellular morphology and tissue architecture to meso-regional organization. RadioDCCD integrated complementary macroscopic information by combining whole-tumor and its habitat-subregions radiomics with a 2D maximal-section heterogeneity metric. These nested models enabled integrated molecular subtype prediction and clinical risk stratification. Across five cohorts totaling 1,659 patients, PathoDCCD reliably recapitulated molecular subtypes, while RadioDCCD provided reliable preoperative prediction. The consistent predictions identified patients with the poorest clinical outcomes. This cross-scale paradigm unifies molecular biology, computational pathology, and quantitative radiology into a biologically grounded strategy for preoperative noninvasive molecular phenotyping of ccRCC.
\end{abstract}

\sloppy

\section*{Introduction}
Clear cell renal cell carcinoma (ccRCC), the most common subtype of renal cancer, exhibits pronounced biological and clinical heterogeneity that contributes to substantial variability in patient outcomes \cite{2017ajcc}. Although TNM staging and pathological features remain central to current clinical decision-making, these metrics incompletely reflect the metabolic and spatial complexity of ccRCC \cite{2017ajcc}. As a result, a subset of patients categorized as early or intermediate stage still experience disease recurrence or progression, underscoring persistent limitations in accurately assessing tumor aggressiveness before surgery \cite{JAMA2024renal,Pre-Metastaticniche2025}.

Molecular profiling has substantially advanced the understanding of ccRCC biology, particularly the metabolic rewiring that underlies tumor progression and therapeutic vulnerability. In particular, molecular analyzes have identified a lipid-deficient, de-clear-cell-differentiated (DCCD) subtype, characterized by dedifferentiated morphology, metabolic reprogramming, and poor prognosis, even in patients who would otherwise be classified as low-risk by conventional clinicopathologic criteria \cite{Multi-omic2024}. Despite its prognostic relevance, molecular testing is not routinely integrated into clinical workflows. Current international guidelines, including those issued by the European Association of Urology (EAU), emphasize pathological parameters such as tumor stage, grade, and the presence of necrosis when guiding recommendations for adjuvant therapy, and do not require molecular or genomic assays for postoperative risk stratification \cite{EAU2024renal,ESMO2024renal,NCCN2025Kidney}. As a result, genomic testing remains rarely applied, limited by cost, specialized resources, and potential delays in initiating treatment. These limitations underscore the need for scalable, noninvasive, and clinically deployable strategies capable of approximating molecular tumor behavior without sequencing.

Histopathology provides complementary information on tumor morphology, captures cellular architecture, stromal composition, and tissue-level organization that reflect metabolic and phenotypic states \cite{ComputationalPathology2015}. Nevertheless, intratumoral heterogeneity imposes substantial barriers. Spatial multi-omics studies reveal that lipid-rich and DCCD regions frequently coexist within individual tumors, forming dynamic dedifferentiation trajectories with distinct metabolic programs. Such compartmental transitions can be subtle and challenging in visual recognition, and needle biopsy sampling often underrepresents aggressive subregions. Existing computational pathology approaches, which typically operate at the patch level, often fail to preserve multi-scale tissue architecture, constraining their ability to detect localized phenotypic shifts within global tissue organization \cite{CLAM2021,HIPT2022,ccRCCdfs2024,pRCCrecurrence2024multi}.

In the preoperative clinical setting, cross-sectional imaging, particularly contrast-enhanced computed tomography (CT), offers the most accessible modality for evaluating renal masses. Radiomics and deep learning studies have demonstrated correlations between CT-derived features and tumor grade, necrosis, stromal composition, and postoperative outcomes \cite{radiomics2012,deeplearning2017}. However, conventional imaging pipelines are still limited in resolving the intratumoral metabolic heterogeneity characteristic of ccRCC. Most approaches rely on handcrafted features, single‑scale analysis, or biologically unconstrained models, which restricts their ability to distinguish lipid‑rich from lipid‑poor regions or to identify molecularly defined subtypes such as DCCD.

Recent advances in artificial intelligence offer a new opportunity to overcome these barriers. In computational pathology, growing evidence shows that transformer-based and multi-scale deep learning architectures, including foundation model style whole slide imaging (WSI) classifiers, significantly improve robustness and scalability across tumor types \cite{GigaPath2024, ROAM2024transformer}. In addition, multimodal deep learning frameworks that fuse histology with genomics or radiology demonstrate superior performance over unimodal models in survival prediction, subtype classification, and prognostic stratification \cite{multimodalccrcc2025multimodal}. These models harness cross-modal supervision to align biological, histological, and radiological representations, offering a biologically grounded route for noninvasive molecular phenotyping. Theoretical and practical advances in self-supervised learning, weak supervision, multi-scale feature extraction, and cross-modal transformer architectures further support the feasibility of such cross-scale integration.

\begin{figure}[htpb!]
  \centering
         \includegraphics[width=0.93\linewidth]{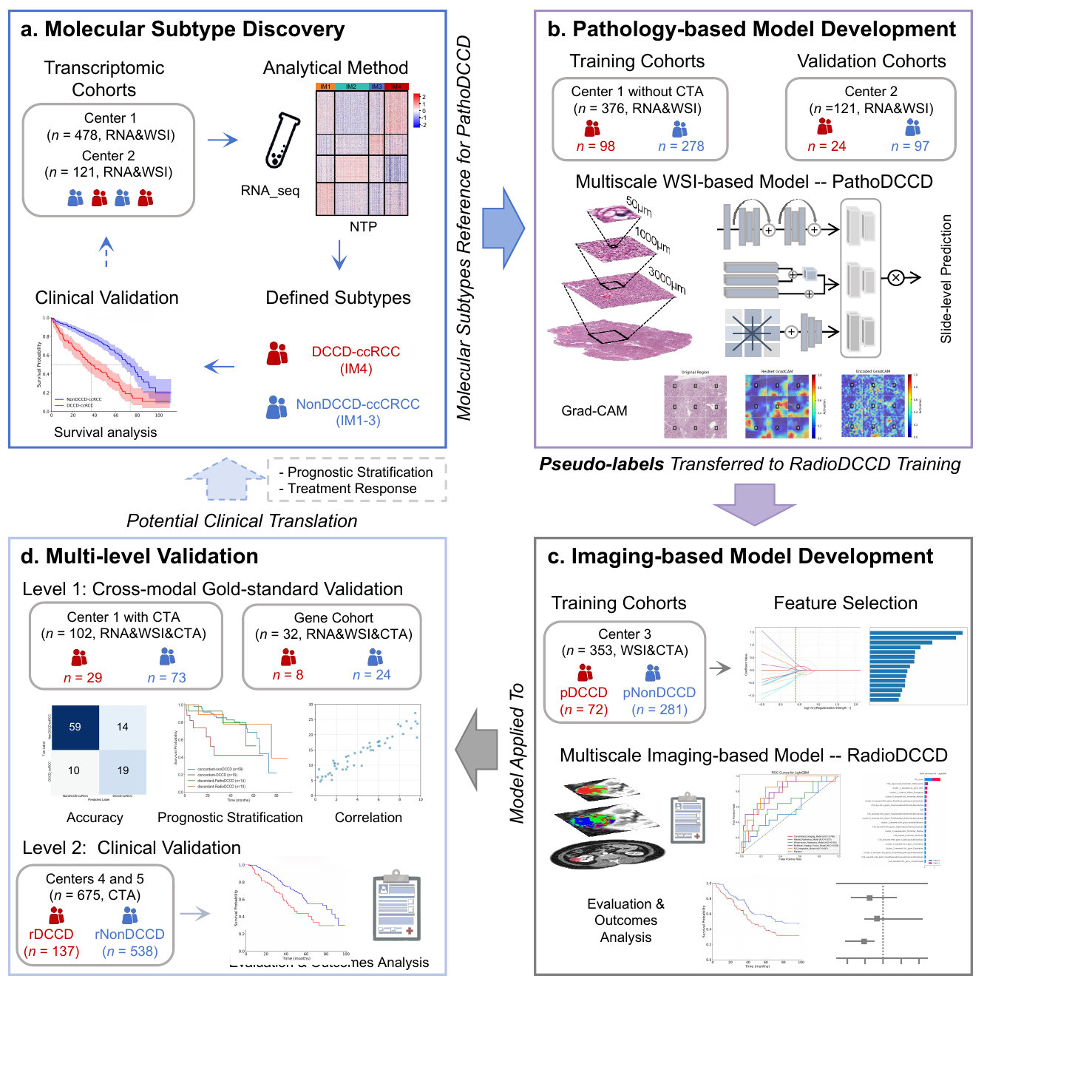}
  \caption{Workflow for molecular subtype discovery and multimodal model development. (a) Molecular gold standard definition. Bulk RNA-sequencing data from two transcriptomic cohorts with paired RNA–WSI (Center 1, $n$ = 478; Center 2, $n$ = 121) were integrated for subtype discovery using Nearest Template Prediction, defining DCCD-ccRCC (IM4) and NonDCCD-ccRCC (IM1–3). This signature was validated through survival and transcriptomic analyses and used as the molecular gold standard. (b) Pathology-based model (PathoDCCD). WSIs from Center 1 without CTA ($n$ = 376) were used for model training, and WSIs from Center 2 ($n$ = 121) for validation, supervised solely by the molecular gold-standard labels. The resulting predictions (pDCCD/pNonDCCD) were used as pseudo-labels for imaging model development. (c) Imaging-based model (RadioDCCD). The radiomics model was subsequently trained on a larger WSI–CTA cohort from Center 3 ($n$ = 353) under supervision from the PathoDCCD pseudo-labels.
(d) Multi-level validation. Cross-modal gold-standard validation was performed on a subset from Center 1 with complete RNA–WSI–CTA data ($n$ = 102) and an independent Gene cohort from Center 3 also with complete RNA–WSI–CTA data ($n$ = 32). External clinical validation of RadioDCCD was performed in two additional CTA-only cohorts (Centers 4–5, $n$ = 675) using predicted subtypes and associated outcome analyses. \\
}\label{fig:workflow}
\end{figure}

In this study, we present a hierarchical cross-scale framework for the preoperative and noninvasive identification of lipid-deficient DCCD-ccRCC (Fig.~\ref{fig:workflow}). The framework captures nested histopathological patterns across the cellular, patch-level, and meso-regional scales through PathoDCCD, while RadioDCCD integrates whole-tumor radiomics, habitat-level subregional descriptors, and maximal-section heterogeneity on CT. molecular supervision transfers molecular signals into both models, allowing predictions to reflect biologically significant variation. Evaluated in five independent cohorts totaling 1,659 patients, the framework accurately recovered molecular subtypes and achieved robust preoperative prediction using CT alone, with consistent cross-modal predictions identifying a distinct high-risk group with markedly poorer outcomes. Building on these task-specific results, this work advances a scalable and biologically informed strategy for the preoperative phenotyping of aggressive ccRCC. The framework brings cellular morphology, tissue architecture, radiological phenotype, and molecular programs into a shared representational space, demonstrating that latent molecular behavior can be inferred from routine clinical data when biological supervision is propagated across modalities. This unified design provides an adaptable foundation for intelligent phenotyping systems that can be extended to other tumor types, molecular states, and clinical decision pathways, expanding the reach of molecularly informed oncology to settings where sequencing is limited and enabling a more accessible and biologically grounded model of precision cancer care.

\section*{Results}
\subsection*{Characteristics of the Study Cohorts}
\begin{figure}[htpb!]
  \centering
         \includegraphics[width=0.93\linewidth]{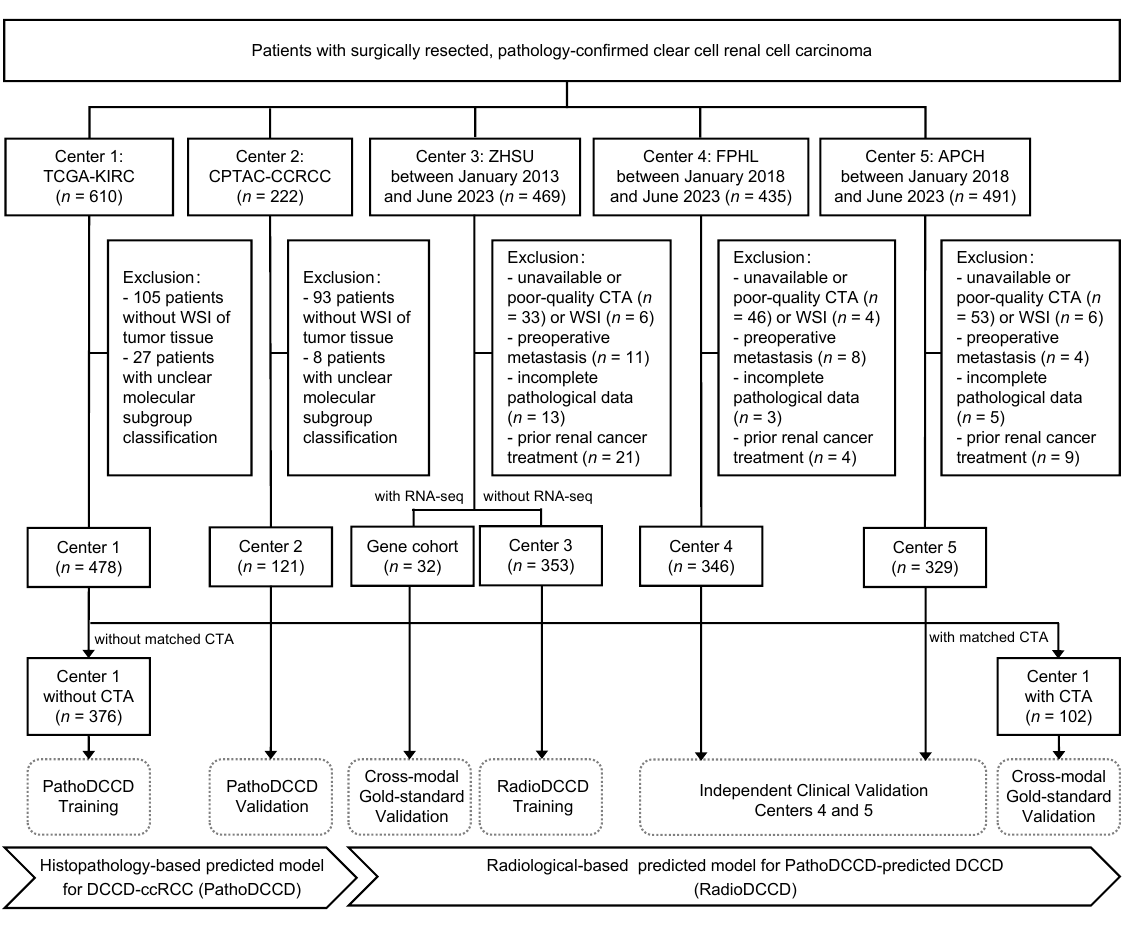}
  \caption{Flowchart of patient cohort selection for multiscale model development and validation. Patients from five independent cohorts with surgically resected ccRCC were selected based on data availability and clinical criteria. The DCCD-ccRCC molecular subtype was first defined using RNA-seq data from Centers 1 and 2 to train the PathoDCCD histopathology model on their WSIs. Using PathoDCCD predictions as pseudo-labels, the RadioDCCD model was subsequently trained on CTA images from the independent Center 3. The RadioDCCD model was then validated against the molecular gold standard in the independent Center 1 with CTA subset, and an independent Gene cohort from Center 3, while the prognostic utility of the PathoDCCD model was further validated in the external clinical cohorts Centers 4 and 5. \\ }
  \label{fig.02}
\end{figure}

The study comprised five independent ccRCC cohorts totaling 1,659 patients, with detailed patient selection and cohort flow depicted in  Fig.~\ref{fig.02}. Demographic and baseline clinical characteristics are summarized in Table~\ref{Table:01}. In the TCGA cohort (Center 1, $n = 478$), patients had a median age of 62 years (range 26–86), with 309 males (64.6\%). The CPTAC-CCRCC cohort (Center 2, $n = 121$) had a median age of 61 years (30–90) and a male predominance of 76.9\%. The ZHSU cohort (Center 3, $n = 353$) presented a median age of 58 years (19–88) with 75.1\% males. The FPHL (Center 4, $n = 346$) and APCH (Center 5, $n = 329$) cohorts showed similar age distributions (median 59 and 60 years) and male proportions (70.8\% and 72.9\%). Tumor sizes were broadly comparable across cohorts, laterality was balanced, and ISUP grade 2–3 and T1 stage predominated. The median follow-up time ranged from 66 to 90 months across the five clinical centers, indicating adequate follow-up for survival analyses. 17.8–38.1\% of patients experienced recurrence, metastasis, or death, with a median disease-free survival (DFS) of approximately 63–82 months across cohorts.

The five cohorts provided complementary cross-modal data essential for model development. Centers 1–2 contributed patients with paired WSIs and RNA sequence profiles, enabling molecular characterization of DCCD biology. PathoDCCD, the histopathology-based DCCD predictor, was trained and internally validated using Center 1 patients without contrast-enhanced arterial-phase CT (CTA) imaging ($n = 376$), and externally validated in Center 2. Center 3 (ZHSU; $n = 353$) then served as the derivation cohort for RadioDCCD, using PathoDCCD-derived pseudo-labels as reference. RadioDCCD was subsequently validated in Center 1 with CTA ($n = 102$) and in the combined Center 4 (FPHL; $n = 346$) and Center 5 (APCH; $n = 329$) cohorts. An independent subset gene cohort from Center 3 ($n = 32$; gene cohort) further supported molecular validation.

\subsection*{Molecular and Clinicopathological Characterization of DCCD-ccRCC}
Molecular profiling in Centers 1 and 2 consistently stratified tumors into DCCD-ccRCC and NonDCCD-ccRCC (Supplementary Figs.1a,2a). KEGG pathway enrichment analysis of the differentially expressed genes revealed significant alterations in pathways regulating lipid and energy homeostasis, including fatty acid degradation and metabolism, glycerolipid metabolism, and steroid biosynthesis. Significant enrichment was also observed in glycolysis, gluconeogenesis and glutathione metabolism, with fatty acid degradation emerging as the most prominently enriched pathway (Supplementary Figs.1b-c,2b-c). Kaplan–Meier curves demonstrated significant differences in both DFS and overall survival (OS) between the two groups ($P <$ 0.001, Supplementary Figs.1d,2d). 
DCCD-ccRCC was associated with older age (median 65 vs.\ 59 years in the combined Centers 1--2 cohort), a higher proportion of males (76.2\% vs.\ 62.5\%), and enrichment of high-grade disease. In Centers 1--2, ISUP grade 3--4 occurred in 51.6\% of DCCD versus 26.4\% in NonDCCD tumors ($P = 0.002$), and T stage $\geq$2 was more frequent in DCCD tumors (27.2\% vs.\ 17.4\%, $P = 0.004$) (Supplementary Figs.1e-h,2e-h, Supplementary Table 1).

\subsection*{Diagnostic Validation of PathoDCCD for WSI-based Prediction of Molecular DCCD}

A hierarchical, multi-branch histopathology model (PathoDCCD) was developed to predict molecular DCCD status directly from diagnostic H\&E WSIs (Fig.~\ref{fig.03}). 
Training used Center 1 (TCGA) patients without CTA ($n$ = 376) with molecular NTP labels as supervision; external validation employed the CPTAC cohort (Center 2, $n$ = 121). In internal validation, PathoDCCD achieved an area under the ROC curve (AUC) of 0.94, accuracy 91\%, sensitivity 0.90, and specificity 0.92. External validation yielded AUC 0.91, accuracy 89\%, sensitivity 0.87, and specificity 0.90, indicating robust generalization across independent data. Hierarchical, multi-region fusion improved slide-level accuracy from 82.5\% (single-scale/naïve aggregation) to 91.3\% (hierarchical fusion). Ablation experiments attributed a 2.6\% absolute AUC gain to the combination of multi-scale structural fusion and graph-based region interaction modules, confirming the value of explicit spatial aggregation. A systematic ablation study confirmed that the deep visual semantic branch constitutes the performance foundation, as its exclusion resulted in a precipitous drop in slide-level AUC to 0.81. Furthermore, replacing graph-based spatial aggregation with naïve concatenation significantly degraded performance (AUC 0.87 vs. 0.94), underscoring the necessity of topological modeling to capture intratumoral heterogeneity. Comprehensive baseline comparisons against standard CNNs and foundation models (e.g., DINOv2), along with detailed ablation metrics, are provided in the Supplementary Information (see Supplementary Table 8).

Interpretability analyses showed that attention heatmaps consistently localized to biologically plausible regions: lipid-depleted areas, necrotic foci, and regions with increased stromal/immune infiltration. Quantitatively, attention hotspots overlapped pathologist annotations in 86.4\% of sampled slides. A downstream confidence-scoring module, designed to downweight low-certainty regions, increased concordance with molecular labels from 78.6\% to 91.2\%, improving robustness to heterogeneous sampling.

% Placeholder: (slide-level calibration plots, decision curve analysis, and per-region ROC curves to be inserted when finalized).

\subsection*{Distribution of PathoDCCD-predicted DCCD and Associated Clinical-Pathological Features} 
Predictions generated by the PathoDCCD model were used as pseudo-labels for imaging model development and were applied to the independent ZHSU cohort (Center 3, $n$ = 353). 
Among 353 pathologically confirmed ccRCC patients in Center 3, 72 (20.4\%) patients were classified as pDCCD (PathoDCCD-predicted DCCD) and 281 (79.6\%) patients as pNonDCCD (PathoDCCD-predicted NonDCCD). Kaplan–Meier curves further demonstrated significant differences in both DFS and OS between the two groups ($P<$ 0.001, Fig.4a-b). Baseline demographics were largely comparable, with a trend toward older age in the pDCCD group (median 61 vs. 58 years, $P$ = 0.089), and no significant differences in sex, BMI, symptoms, or comorbidities (all $P>$  0.05; Supplementary Table 2). Pathologically, pDCCD tumors exhibited more aggressive features, including larger size (6.8 vs. 6.0 cm, $P$ = 0.002), higher rates of necrosis (38.9\% vs. 15.7\%, $P <$ 0.001) and sarcomatoid differentiation (8.3\% vs. 1.4\%, $P$ = 0.008), higher ISUP grade (grade 3–4: 55.6\% vs. 27.8\%, $P<$ 0.001), and more advanced T stage (T3–4: 26.4\% vs. 9.2\%, $P <$ 0.001) (Fig.4g-j). Consistently, clinical outcomes were worse: median disease-free survival was 41 vs. 78 months ($P<$  0.001), with 44.4\% of pDCCD patients experiencing recurrence, metastasis, or death compared with 16.7\% of pNonDCCD patients ($P<$ 0.001). 

Univariable logistic regression identified larger tumor size, higher ISUP grade, advanced T stage, prominent arterial-phase enhancement, higher RENAL score, and greater intratumoral heterogeneity as significant predictors of pDCCD (Supplementary Table 3). In multivariable analysis adjusting for age, ISUP grade, and T stage, independent pathological predictors were ISUP grade 4 (adjusted OR = 5.21, 95\% CI: 1.42–19.12, $P$ = 0.013) and advanced T stage (T3: OR = 3.86, 95\% CI: 1.55–9.62, $P$ = 0.004; T4: OR = 8.45, 95\% CI: 1.82–39.24, $P$ = 0.007). Independent imaging predictors included arterial-phase enhancement (OR = 2.12, 95\% CI: 1.21–3.72, $P$ = 0.009), higher RENAL score (OR = 1.18, 95\% CI: 1.01–1.38, $P$ = 0.034), and increased intratumoral heterogeneity (OR = 2.40, 95\% CI: 1.30–4.43, $P$ = 0.005). Tumor size was no longer significant after adjustment, suggesting its effect is captured by T stage and composite imaging metrics.

\subsection*{Diagnostic Validation of RadioDCCD for Preoperative  Imaging-based Prediction of PathoDCCD-predicted DCCD}
The CT-based radiomics framework, RadioDCCD, was developed for preoperative identification of DCCD-ccRCC using PathoDCCD-derived pseudo-labels. Three types of imaging features were included: a) six conventional imaging features, including maximum tumor diameter, degree of arterial-phase enhancement, neovascularization score, RENAL score, renal contour irregularity, and intratumoral heterogeneity (ITH) score; b) habitat-based radiomics features, defined as the combination of features from three intratumoral habitats (2,553 features); and c) whole-tumor radiomics features (1,687 features). Specific feature descriptions and the feature selection process are provided in Supplementary Methods.

Five modeling strategies were evaluated using four machine learning algorithms (LightGBM, XGBoost, Logistic Regression, and a Stacking Ensemble). These strategies comprised: (a) a conventional imaging model; (b) a habitat radiomics model; (c) a whole-tumor radiomics model; (d) a multilevel imaging fusion model; and (e) a full integrative model, which combined multilevel imaging features with clinical variables including age, sex, ISUP grade, and T stage. All models were trained and evaluated under identical conditions, with performance assessed on a held-out test set comprising 20\% of the data (random state = 42) using the AUC.

Among the algorithms, tree-based ensembles demonstrated the strongest performance with complex feature sets. The LightGBM algorithm paired with the full integrative model achieved the highest overall discrimination, yielding an AUC of 0.847. XGBoost performed comparably on the same feature set (AUC = 0.837). Notably, both algorithms also showed robust performance with the multilevel imaging fusion model (LightGBM: 0.828; XGBoost: 0.835). In contrast, Logistic Regression performed best with the conventional imaging model (AUC = 0.827), outperforming its results with more complex feature combinations. The Stacking Ensemble did not yield a consistent performance gain; its best result was with the conventional imaging model (AUC = 0.815), and it underperformed relative to LightGBM and XGBoost on the integrative model (AUC = 0.779).

Across the five feature strategies, the full integrative model consistently delivered the highest discrimination when paired with tree-based methods. The multilevel imaging fusion model also showed strong predictive capability with these algorithms. The conventional imaging model proved to be a reliable and well-performing feature set across all classifiers, particularly with Logistic Regression. Conversely, the habitat radiomics model exhibited the weakest discriminative power (AUC range: 0.380–0.598), while the whole-tumor radiomics model showed intermediate but generally inferior performance compared to the fusion-based and integrative approaches (Fig.5a, Supplementary Table 4).

Performance evaluation of machine learning models across different feature sets were shown in Supplementary Fig.4a. SHAP-based interpretability analyses highlighted high-order wavelet textures, arterial-phase enhancement heterogeneity, and marginal-zone contrast patterns as key discriminative traits in DCCD-predicted tumors (Supplementary Fig.4b).

External validation of RadioDCCD was conducted at cross-modal and clinical levels. In the Center 1 with CTA cohort ($n$ = 102), comparison with molecular ground-truth labels demonstrated an accuracy of 76.5\%, with a sensitivity of 65.5\%, specificity of 80.8\%, and an AUC of 0.927 (Fig.5b–c), indicating robust imaging-based discrimination of DCCD status (Supplementary Tables 5 and 6).

Validation was further extended to an independent gene-validated cohort from Center 3 ($n$ = 32), in which RNA sequencing enabled direct cross-modal comparison (Fig.5d–e). High concordance was observed between pathology-based predictions and molecular labels (accuracy 90.6\%, AUC = 0.99). RadioDCCD showed strong agreement with PathoDCCD (accuracy 84.4\%, AUC = 0.97) and retained high discriminative performance against molecular labels (accuracy 81.2\%, AUC 0.96), supporting its cross-modal generalizability. Multi-modal interpretability analyses of deep learning models for discriminating DCCD-ccRCC and NonDCCD-ccRCC subtypes were presented in Fig.6.

Clinical validation was performed in two independent CTA-only cohorts (Centers 4 and 5). Patients stratified by RadioDCCD into rDCCD (RadioDCCD-predicted DCCD) and rNonDCCD (RadioDCCD-predicted NonDCCD) groups exhibited significantly different survival outcomes in both cohorts (both $P <$ 0.001; Fig.4c–f), confirming the prognostic relevance of RadioDCCD in fully preoperative settings.

\subsection*{Multi-Scale, Cross-Modal Models Improve Prognostic Stratification}
To investigate the complementary prognostic value of PathoDCCD and RadioDCCD, multi-scale cross-modal integration was performed to refine risk stratification in ccRCC. Patients from the Center 1 with CTA cohort ($n$ = 102), with both PathoDCCD predictions and preoperative CT-based RadioDCCD assessments available, were stratified into four subgroups based on prediction concordance: concordant-DCCD (both modalities predicting DCCD), concordant-nonDCCD (both predicting NonDCCD), discordant-PathoDCCD (PathoDCCD-predicted DCCD with RadioDCCD NonDCCD), and discordant-RadioDCCD (PathoDCCD NonDCCD with RadioDCCD-predicted DCCD).

Kaplan–Meier analyses demonstrated distinct survival profiles across these groups. The concordant-nonDCCD group was the largest ($n$ = 59) and showed a relatively favorable prognosis, with an event rate of 27.1\% and a median survival of 65.3 months. In contrast, the concordant-DCCD group ($n$ = 19) exhibited the poorest outcomes, characterized by the highest event rate (36.8\%) and the shortest median survival (25.7 months). Among discordant cases, the discordant-PathoDCCD group ($n$ = 14) had the lowest event rate (21.4\%), and median survival was not reached during follow-up, indicating comparatively favorable outcomes. The discordant-RadioDCCD group ($n$ = 10) displayed an intermediate risk profile, with an event rate of 30.0\% and a median survival of 75.6 months.

\section*{Discussion}
In this multi-institutional study, we developed and validated a hierarchical AI framework that bridges molecular subtyping, histopathology, and preoperative imaging to identify a clinically aggressive, lipid-deficient subtype of ccRCC. Our findings confirm that DCCD-ccRCC is associated with a significantly worse prognosis, independent of conventional staging. While previous studies have established the prognostic value of molecular subtyping in ccRCC \cite{Multi-omic2024}, the clinical translation has been hampered by the dependency on costly and tissue-consuming genomic assays. Here, we show that this molecular phenotype can be accurately inferred from standard H\&E-stained slides and further predicted non-invasively using preoperative CTA scans. This cross-modal translation establishes a practical and scalable strategy to make molecular stratification accessible in routine clinical workflows.

The robust performance of our PathoDCCD model stems from its multi-scale architecture, which is specifically designed to address the profound intratumoral heterogeneity of ccRCC. While prior computational pathology models have successfully predicted grade, stage, or mutations from WSIs \cite{WSIsgao2024deep, WSIs2jiang2025diagnosis, WSIs3mao2025rpf, WSIs4chen2024deep}, they often rely on global feature aggregation and may overlook critical spatial relationships. In contrast, our approach, integrating cellular composition, patch-level semantics, and spatial context through graph-based aggregation, explicitly captures the subtle morphological correlates of lipid depletion. This design is conceptually aligned with the biological understanding of DCCD-ccRCC as a spatially heterogeneous entity, enabling our model to identify features that are challenging for the human eye to quantify consistently. The high concordance between model attention and pathologist annotations further advances the field by providing not just prediction but also interpretable, biologically grounded insights.

Extending this stratification to the preoperative phase, our RadiopDCCD model successfully linked CT-derived radiomic features to the DCCD phenotype. The enrichment of features related to arterial enhancement and intratumoral heterogeneity aligns with the known hypervascular and morphologically complex nature of aggressive RCC \cite{RCC1udayakumar2021deciphering, RCC2zhou2023spatio, RCC3ran2020low, RCC4tabata2023inter}. However, moving beyond conventional radiological semantics, our radiomics approach quantifies these phenotypes in an objective and reproducible manner. Previous radiomics studies have largely focused on predicting survival or Fuhrman grade \cite{RADIOMICS1raman2024radiomics, RADIOMICS2wang2020differentiation, radiomics3gouravani2025diagnostic, radiomics4he2025ai}. Our model's novelty lies in its direct linkage to a well-defined molecular subtype, creating a non-invasive bridge to underlying tumor biology. The robust performance of the radiomics-only model, with limited incremental value from standard clinical variables, underscores that imaging features can capture intrinsic biology beyond conventional parameters, echoing the findings of other studies in oncology that highlight the superior prognostic value of deep phenotypes over clinical covariates \cite{phenotypes1flynn2023deep, phenotypes2reicher2025deep, phenotypes3an2023deep}.

The observed discordance between radiological and molecular subtyping in a subset of patients offers critical insights. While technical factors contributed, biological and methodological reasons are plausible. This discrepancy underscores a fundamental challenge in radiophenomics: the fidelity with which macroscopic imaging can reflect molecular alterations. Our findings suggest that for some tumors, the molecular signature may not be fully manifested in the imaging phenotype, a phenomenon observed in other cancers where genomic-intrinsic imaging subtypes show incomplete overlap \cite{genomic1mendes2023current, genomic2tian2025genome}. Furthermore, our multi-region pathology analysis reveals a potential limitation of whole-tumor radiomics: the signal from critical but geographically limited DCCD foci may be diluted, a challenge less pertinent to models predicting more globally expressed features like mutations. Future studies with spatially matched multi-omics and imaging data are needed to dissect this complex relationship.

Our study has several limitations. The reliance on histopathology-derived pseudo-labels for radiomics model development, while a pragmatic solution, means the ground truth for external CT cohorts is inferred. Furthermore, institutional variations in imaging protocols, despite preprocessing normalization, pose challenges to broad generalizability. The retrospective design also necessitates prospective validation to confirm the clinical utility of our models.

In summary, we have established a coherent pipeline from molecular definition to non-invasive identification of a high-risk ccRCC subtype. The PathoDCCD and RadiopDCCD models provide complementary tools for pathological refinement and preoperative risk assessment, respectively. A promising clinical application lies in identifying early-stage patients with a DCCD phenotype who, despite favorable conventional staging, might be candidates for more intensive surveillance or enrollment into trials for adjuvant therapies. By making molecular subtyping accessible through routine clinical data, our framework represents a significant step towards personalized management of ccRCC.

\section*{Methods}

\subsection*{Ethics statement}
This study complies with all relevant ethical regulations and was approved by the Institutional Review Boards of the Zhongda Hospital Affiliated to Southeast University. The informed consent was waived because patients were not directly recruited for this study. Data from TCGA-KIRC and CPTAC-CCRCC was open-accessed, ensuring patient anonymity without risk of patient identification.

\subsection*{Study Design and Cohorts}
The overall study workflow is summarized in Fig. \ref{fig:workflow}.

\paragraph{Center 1: TCGA Cohort} \;  
The TCGA-KIRC dataset was used for pathological model development (PathoDCCD) and initially included 610 ccRCC patients. After excluding 73 cases without diagnostic tumor tissue WSIs, 537 patients remained with RNA sequencing, diagnostic H\&E WSIs, and clinical information. Patients without molecular subtyping by nearest template prediction (NTP) were further excluded, leaving 478 patients with complete molecular, histopathological, and follow-up data (331 alive, 147 deceased). Abdominal CTA was available for 102 patients. The cohort was thus stratified into Center~1 without CTA ($n = 376$) and Center~1 with CTA ($n = 102$).

\paragraph{Center 2: CPTAC-CCRCC Cohort}  \;
The CPTAC-CCRCC dataset was used for external validation of the PathoDCCD model and initially included 222 patients with ccRCC. After excluding 93 cases without diagnostic tumor tissue WSIs, 129 patients remained with available RNA sequencing, diagnostic H\&E WSIs, and clinical information. Further exclusion of patients without molecular subtyping by NTP resulted in 121 patients with complete molecular, histopathological, and follow-up data (97 alive, 24 deceased).

\paragraph{Center 3: ZHSU Cohort}  \;
This study retrospectively enrolled 437 patients with ccRCC who underwent surgical resection at Zhongda Hospital, Southeast University (ZHSU) between January 2013 and January 2023. Patients were excluded for missing or poor-quality CTA images ($n = 33$), missing or poor-quality diagnostic WSIs ($n = 6$), preoperative distant metastasis ($n = 11$), incomplete pathological data ($n = 13$), or a prior history of renal cancer treatment ($n = 21$). The final cohort included 353 patients with available CTA images and diagnostic H\&E WSIs. Patients were stratified into DCCD-pDCCD (PathoDCCD-predicted de-clear cell differentiated ccRCC) and NonDCCD-pDCCD (PathoDCCD-predicted non-de-clear cell differentiated ccRCC) groups based on PathoDCCD predictions.

\paragraph{Center 4: FPHL Cohort}  \;
A total of 411 patients with ccRCC who underwent surgery at the First People’s Hospital of Lianyungang between January~2019 and January~2023 were retrospectively reviewed. After excluding those with missing or poor-quality CTA ($n = 46$), preoperative distant metastasis ($n = 12$), incomplete pathology ($n = 3$), or prior renal cancer treatment ($n = 4$), 346 patients with available CTA were retained.

\paragraph{Center 5: APCH Cohort}  \;
A total of 421 patients with ccRCC who underwent surgery at Anhui Provincial Cancer Hospital between January 2019 and January 2023 were retrospectively reviewed. After excluding patients with missing or poor-quality CTA ($n = 53$), preoperative distant metastasis ($n = 19$), incomplete pathology ($n = 6$), or prior renal cancer treatment ($n = 14$), 329 patients with available CTA were retained.

\paragraph{Gene Cohort} \;
FFPE tumor blocks from 32 ccRCC patients who underwent surgical resection at Zhongda Hospital, Southeast University (Center 3), between January 2019 and January 2020 were collected. RNA was extracted from these tissue blocks and subjected to RNA sequencing. Matched diagnostic H\&E WSIs, CTA, and clinical follow-up data were also obtained. Sequencing library type, platform, read length, and sequencing depth were recorded for all samples.

\subsection*{Molecular Subtyping}

Immune-related gene signatures reported in \cite{Multi-omic2024} were used to classify patients into DCCD and NonDCCD subtypes of ccRCC. Signature gene templates were applied to log$_2$-transformed and RLE-normalized RNA-seq matrices using the nearest template prediction (NTP) algorithm implemented in the \texttt{CMScaller} package (\texttt{nPerm = 1000}, \texttt{seed = 42}, \texttt{nCores = 30}). Center~1 (TCGA-KIRC) and Center~2 (CPTAC-CCRCC) cohorts were analyzed independently, and patients were assigned to molecular subtypes based on robust template matching.

\subsection*{PathoDCCD Model: Methodology for Histological Feature Encoding and Multiscale Classification}

\begin{figure}[htpb!]
  \centering
         \includegraphics[width=0.93\linewidth]{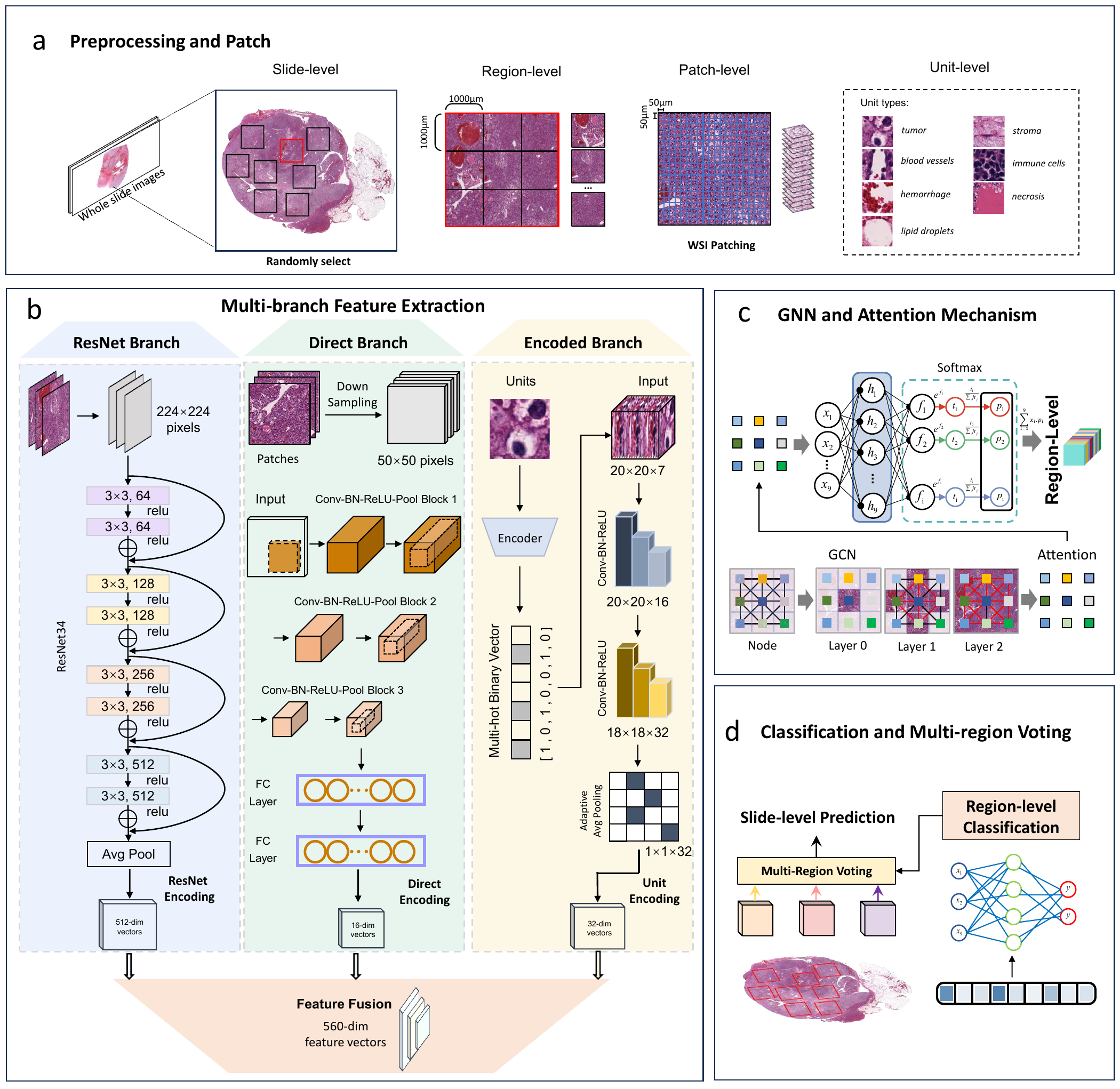}
  \caption{Overview of the PathoDCCD multi-branch, multi-scale deep learning network for DCCD-ccRCC classification. The network integrates histological features across multiple scales: micro-unit (50 × 50 µm), patch (1000 × 1000 µm), region (3000 × 3000 µm), and WSI-level (global context). Cellular-level features from seven tissue components are extracted using a ResNet34-based micro-unit branch, while patch-level high-level visual features and color-texture features are obtained via ResNet34 and a lightweight CNN branch, respectively. The resulting feature vectors are concatenated into a 560-dimensional representation for downstream analysis. Spatial relationships between adjacent patches are modeled using an attention-based GNN, and multi-region voting (hard, soft, and confidence-weighted) is applied to integrate predictions across multiple regions per slide. The network is trained in a multi-stage procedure, including micro-unit pre-training, branch-wise training, and end-to-end fine-tuning with ResNet34 frozen. Attention-GNN and multi-region aggregation enhance robustness and accuracy in distinguishing DCCD-ccRCC from NonDCCD-ccRCC regions. \\
\textbf{Abbreviations}: DCCD-ccRCC: de-clear cell differentiated clear cell renal cell carcinoma; NonDCCD-ccRCC: non-de-clear cell differentiated clear cell renal cell carcinoma; Conv-BN-ReLU: convolution + batch normalization + rectified linear unit; FC: fully connected layer; Avg Pool: average pool; GNN: graph neural network; GCN: graph convolutional network.
}\label{fig.03}
\end{figure}

PathoDCCD is a hierarchical, multi-branch model that integrates unit-, patch-, and region-level histological information to classify DCCD-ccRCC from H\&E WSIs. The workflow corresponds to Fig.~\ref{fig.03}a--d: (a) multi-branch feature extraction, (b) graph-based spatial modeling with attention, (c) region-level classification, and (d) multi-region voting for slide-level prediction.

\paragraph{Data Preprocessing and Hierarchical Sampling (Fig.~\ref{fig.03}a)} \;
WSIs are processed with a four-tier sampling scheme: WSI-level tissue detection, region-level windows of $3000\times 3000~\mu$m, patch-level tiles of $1000\times 1000~\mu$m (arranged as a $3\times 3$ grid per region), and micro-units of $50\times 50~\mu$m (yielding a $20\times 20$ grid per patch). This design preserves cellular composition, mesoscopic morphology, and local spatial context.

\paragraph{Multi-Branch Feature Extraction and Fusion (Fig.~\ref{fig.03}a)}  \;
Each patch is encoded by three complementary branches: (i) a unit-level tissue encoder that classifies seven tissue components (tumor, stroma, blood vessels, immune cells, hemorrhage, necrosis, lipid droplets) and compresses the resulting $20\times 20\times 7$ tensor into a 32-d vector; (ii) a ResNet34 branch applied to the $224\times 224$ patch for 512-d semantic features; and (iii) a lightweight CNN that summarizes color/texture into a 16-d vector. The concatenation yields a 560-d patch descriptor.

\paragraph{Graph Neural Network and Attention (Fig.~\ref{fig.03}b)}  \; 
Within each region, a patch adjacency graph (9 nodes, 8-neighborhood) is constructed. A two-layer GNN refines patch descriptors by aggregating spatially adjacent context. Gated attention pooling assigns data-driven weights to patches and forms a region-level representation.

\paragraph{Region-Level Classification (Fig.~\ref{fig.03}c)}  \;
A shallow MLP maps the attention-aggregated region vector to a DCCD probability $p_\mathrm{region}$.

% \paragraph{Multi-Region Voting for Slide-Level Prediction (Fig.~\ref{fig.03}d)}  \;
% For each WSI, 8--10 regions are sampled. Slide-level probability is computed by confidence-weighted soft voting:
% \textcolor{red}{
% \begin{equation}
% p_\mathrm{slide}=\frac{\sum_{r} c_r\, p_r}{\sum_{r} c_r},\quad
% c_r = 1-\frac{H(p_r)}{\log 2},\;\; H(p)=-p\log p-(1-p)\log(1-p),
% \end{equation}}
% \textcolor{red}{where....}with hard/soft voting reported for ablations.

\paragraph{Multi-Region Voting for Slide-Level Prediction (Fig.~\ref{fig.03}d)} For each WSI, $8$--$10$ regions are sampled. The final slide-level probability is computed via confidence-weighted soft voting:
\begin{equation}
    p_{\mathrm{slide}} = \frac{\sum_{r} c_r p_r}{\sum_{r} c_r}, \quad 
    c_r = 1 - \frac{H(p_r)}{\log 2},
\end{equation}
where $H(p) = -p \log p - (1-p) \log(1-p)$ is the binary entropy function. In this formulation, $p_r$ denotes the predicted probability of the DCCD subtype for the $r$-th region, and $c_r \in [0, 1]$ represents the confidence weight, which automatically down-weights regions with high predictive uncertainty (entropy) to ensure robust slide-level classification.

\paragraph{Design Rationale and Contributions} \; 
(i) Structured unit-level encoding exposes lipid-deficient morphology via explicit tissue composition. (ii) Complementary patch features (semantic/texture/composition) increase discriminability and interpretability. (iii) Graph-based aggregation captures spatial organization critical for intratumoral heterogeneity. (iv) Confidence-weighted multi-region voting improves slide-level robustness. Full implementation details are provided in Supplementary Information.

\subsection*{Clinical Characteristics} 
Clinical characteristics collected through electronic health records included age, gender, tumor history, smoking history, alcohol consumption history, hypertension history, diabetes history, clinical symptoms (gross hematuria, low back pain, upper abdominal pain), height, weight, body mass index (BMI), serum calcium (mmol/L), serum urea (mmol/L), serum creatinine ($\mu$mol/L), triglyceride (mmol/L), total cholesterol (mmol/L), urine occult blood, urine protein, urine red blood cells (cells/$\mu$L), urine white blood cells (cells/$\mu$L), and history of targeted immunotherapy. 

Pathological information was collected through postoperative pathological reports, including surgical procedure, side of the tumor, location of the tumor, tumor size (cm), histological grade (ISUP/WHO grade), necrosis percentage, TNM stage, and tumor invasion status (renal pelvis and calyces, renal sinus, perirenal fat, renal vein, vena cava, and vasculature).

\subsection*{CT Protocols} 
CT images were acquired in the transverse plane, including noncontrast and arterial, portal venous, and delayed phases postcontrast. Detailed parameters for CT examinations are provided in Supplementary Table 7.

\subsection*{Imaging Features Assessment} 
For patients with multiple lesions, the largest tumor was selected as the main target for ccRCC analysis. Three readers (X.G.P., Y.C., and F.L., with 12-year, 5-year, and 3-year experience in abdominal imaging, respectively) independently and blindly reviewed all images to evaluate the radiologic features. Any disagreements were resolved by majority vote. 

The following six imaging features were evaluated on CTA images:  
(a) maximum tumor diameter,  
(b) degree of arterial-phase enhancement,  
(c) neovascularization score,  
(d) RENAL score,  
(e) renal contour irregularity, and  
(f) intratumoral heterogeneity (ITH) score.  

Details regarding the evaluation of radiologic features are available in Supplementary Information.

\subsection*{RadiopDCCD Model: Methodology for Radiomic-Based Classification} 
\textit{RadiopDCCD}, a radiomics-based model trained with pseudo-labels generated by the \textit{PathoDCCD} classifier, was developed to predict DCCD-pDCCD and NonDCCD-pDCCD tumors. The model was constructed using contrast-enhanced CT scans from Center~3 ($n = 353$).

\paragraph{Pseudo-labels and Data Source} \; 
Pseudo-labels were derived from the \textit{PathoDCCD} model predictions on corresponding WSIs, ensuring consistent DCCD annotation across modalities.

\paragraph{Image Segmentation and Preprocessing}  \;
All tumors were segmented using a deep learning–based nnU-Net model, followed by manual refinement by an abdominal radiologist with five years of experience to ensure accuracy. CTA images and corresponding masks were resampled to isotropic voxel spacing of $1.0 \times 1.0 \times 1.0~\text{mm}^3$ using B-spline interpolation. Voxel intensities were normalized via z-score standardization to reduce inter-scanner variability.

\paragraph{Radiomic Feature Extraction and Selection}  \;
Radiomic features were extracted using the PyRadiomics-based FAE platform, encompassing first-order statistics, shape-based descriptors, and multiple texture matrices derived from original, wavelet, and Laplacian of Gaussian (LoG)–filtered images.  
Feature reduction was performed through a multi-step process:  
(i) removal of low-variance and high-missing-value features,  
(ii) correlation filtering (Pearson $r > 0.90$),  
(iii) univariate analysis using Student’s \textit{t}-test or Mann–Whitney \textit{U} test, and  
(iv) least absolute shrinkage and selection operator (LASSO) logistic regression with ten-fold cross-validation.  
This pipeline yielded a compact subset of the most predictive and non-redundant features.

\paragraph{RadiopDCCD Machine Learning Model Development}  \;
All models were implemented using Python. LightGBM and XGBoost classifiers were utilized from their respective libraries (LightGBM v4.1.0; XGBoost v2.0.0), while Logistic Regression and the base estimators for the Stacking Ensemble were built using scikit-learn (v1.2). To mitigate potential class imbalance, the class\_weight parameter was set to 'balanced' for Logistic Regression, and corresponding balancing techniques were applied for ensemble methods. Hyperparameter tuning was conducted via a five-fold cross-validated grid search on the training set. Key parameters optimized included the number of estimators, learning rate, maximum tree depth, and regularization terms for tree-based models, and the regularization strength for Logistic Regression. The model configuration yielding the highest mean cross-validation AUC was retained as the final model for each algorithm.

\paragraph{Integrated Model Construction}  \;
To evaluate the additive value of clinical information, a Full Integrative Model was developed. This model concatenated the extracted imaging features (from conventional and/or radiomic analyses) with key clinical variables, including age, sex, ISUP grade, and T stage. This combined feature vector was then used to train and test the same suite of machine learning classifiers (LightGBM, XGBoost, Logistic Regression, Stacking Ensemble) under an identical training and evaluation framework, allowing for a direct comparison of predictive performance with imaging-only models.

\paragraph{Model Evaluation and Interpretation}  \;
Both radiomics-only and hybrid models were evaluated using repeated five-fold cross-validation. Discriminative performance was assessed with the area under the ROC curve (AUC), along with accuracy, sensitivity, specificity, and F1 score. Calibration performance was evaluated using calibration curves and the Hosmer–Lemeshow test, while clinical utility was examined using decision curve analysis (DCA).  
Model interpretability was analyzed through SHapley Additive exPlanations (SHAP), providing both global and patient-level insights into feature contributions.  

Detailed implementation procedures and parameter settings are provided in Supplementary Information.

\subsection*{Survival Outcomes} 
DFS was defined as the time from surgery to the first recurrence, progression, or last follow-up for censored cases. OS was defined as the time from surgery to death from any cause or last follow-up.

\subsection*{Statistical Analysis} 
Statistical analysis was performed with Python (version 3.9.0; \url{https://www.python.org/}) and R software (version 4.1.0; \url{https://www.r-project.org/}). Continuous variables were reported as medians and interquartile ranges (IQRs), while categorical variables were summarized as numbers and percentages. Statistical significance was defined as $P<0.05$.

Spearman correlation analysis was used to assess associations between non-normally distributed or ordinal variables. Univariate and multivariable logistic regression analyses were conducted to identify risk factors, with corresponding odds ratios (OR) and 95\% confidence intervals (CI). Model performance comparisons were conducted using the DeLong test and integrated discrimination improvement (IDI). The discriminative ability of models was evaluated with the area under the receiver operating characteristic curve (AUC), complemented by accuracy, sensitivity, specificity, and F1 score.

Calibration plots were generated to compare predicted and observed values, while decision curve analysis (DCA) was applied to assess the clinical net benefit across varying threshold probabilities. Model interpretability was analyzed using SHapley Additive exPlanations (SHAP), providing both global and individual-level feature contributions.

Kaplan--Meier survival analysis was performed using the \texttt{scikit-survival} Python package to evaluate the prognostic significance of the molecular subtype system across two independent cohorts. Statistical significance was determined by the log-rank test, with $P<0.05$ considered statistically significant.

\section*{Data Availability}
The publicly available TCGA-KIRC and CPTAC-CCRCC datasets can be accessed at the Genomic Data Commons portal (\url{https://gdc.cancer.gov/}). Expression matrices of TCGA-KIRC along with clinical features were obtained from UCSC Xena (\url{https://xenabrowser.net/datapages/?cohort=GDC%20TCGA%20Kidney%20Clear%20Cell%20Carcinoma%20(KIRC)&removeHub=https%3A%2F%2Fxena.treehouse.gi.ucsc.edu%3A443}). Expression matrices of CPTAC-CCRCC along with clinical features were obtained from UCSC Xena (\url{https://xenabrowser.net/datapages/?cohort=GDC%20CPTAC-3&removeHub=https%3A%2F%2Fxena.treehouse.gi.ucsc.edu%3A443}). Signature genes of DCCD-ccRCC used in NTP classification were obtained from the supplementary material of the original paper (\url{https://pmc.ncbi.nlm.nih.gov/articles/PMC10937392/#_ad93_}). The KiTS2023 database is available at \url{https://kits-challenge.org/kits23/}.

\section*{Code Availability}
The source code and trained model weights of PathoDCCD Model are publicly available at \url{https://github.com/ContinuumCoder/DeepDCCD-Multiscale.git}. The code for this study is available upon request.

% \begin{figure}[htpb!]
%   \centering
%          \includegraphics[width=0.95\linewidth]{Figures/fig_3.pdf}
%   \caption{Flow diagram of study cohorts and data allocation. This diagram illustrates patient cohorts used for the development and validation of the PathoDCCD pathological model and radiological models. Initial WSI and molecular data were derived from Center 1 (n = 478) and Center 2 (n = 121), with an independent gene cohort (n = 30). PathoDCCD training and internal validation used Center 1 without CTA (n = 376), while Center 2 served as external validation. Center 3 (n = 353) provided PathoDCCD-predicted molecular subtypes, which were used as labels to construct the RadiopDCCD model. RadiopDCCD model validation was performed using Center 1 with CTA (n = 102) and the combined cohorts from Centers 4 and 5 (n = 675). Center 1: TCGA cohort. Center 2: CPTAC-CCRCC cohort. Center 3: ZHSU cohort. Center 4: FPHL cohort. Center 5: APCH cohort.Abbreviations: WSI, whole-slide image; CTA, contrast-enhanced arterial-phase CT; DCCD-ccRCC, de-clear cell differentiated clear cell renal cell carcinoma; NonDCCD-ccRCC, non-de-clear cell differentiated clear cell renal cell carcinoma; DCCD-pDCCD, PathoDCCD-predicted de-clear cell differentiated clear cell renal cell carcinoma; NonDCCD-pDCCD, PathoDCCD-predicted non-de-differentiated clear cell renal cell carcinoma.
% }\label{fig:02}
% \end{figure}

\newpage

\newpage
\begin{figure}[htpb!]
  \centering
         \includegraphics[width=0.98\linewidth]{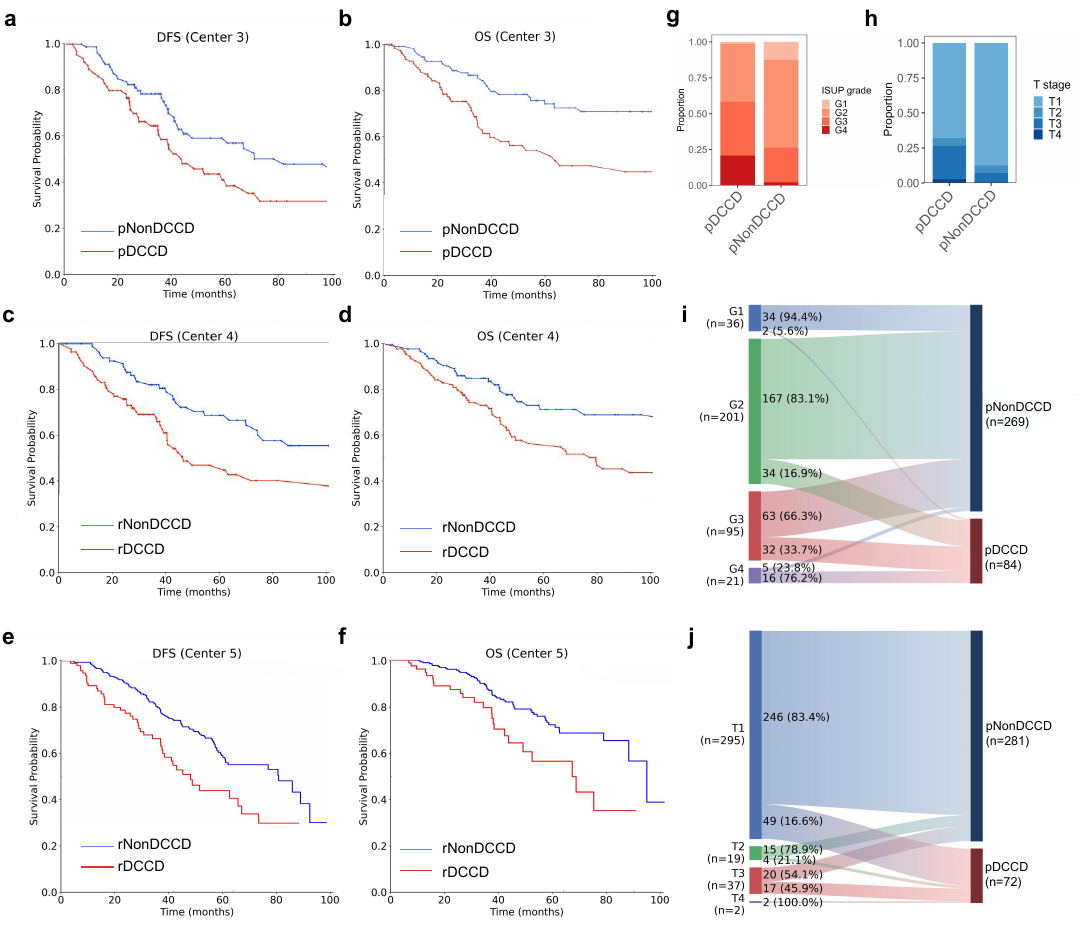}
  \caption{Survival and clinicopathological characteristics of pDCCD and pNonDCCD in Center 3 and rDCCD and rNonDCCD in Centers 4 and 5. (a–b) KM curves for DFS (a) and OS (b) comparing pDCCD versus pNonDCCD in Center 3 (log-rank P $<$0.001). (c–d) DFS (c) and OS (d) KM curves for rDCCD versus rNonDCCD in Center 4 (log-rank P $<$ 0.001). (e–f) DFS (e) and OS (f) KM curves for rDCCD versus rNonDCCD in Center 5 (log-rank P $<$ 0.001). (g–j) Distribution of ISUP grade and T stage between pDCCD and pNonDCCD in Center 3, shown using stacked bar charts (g: ISUP grade; h: T stage) and corresponding Sankey diagrams (i: ISUP grade; j: T stage). \\
\textbf{Abbreviations}: pDCCD: PathoDCCD-predicted de-clear cell differentiated ccRCC; pNonDCCD: PathoDCCD-predicted non-de-clear cell differentiated ccRCC; rDCCD: RadioDCCD-predicted de-clear cell differentiated ccRCC; rNonDCCD: RadioDCCD-predicted non-de-clear cell differentiated ccRCC;KM: Kaplan–Meier; DFS, disease-free survival; OS, overall survival.
}\label{fig.04}
\end{figure}

\newpage
\begin{figure}[htpb!]
  \centering
         \includegraphics[width=\linewidth]{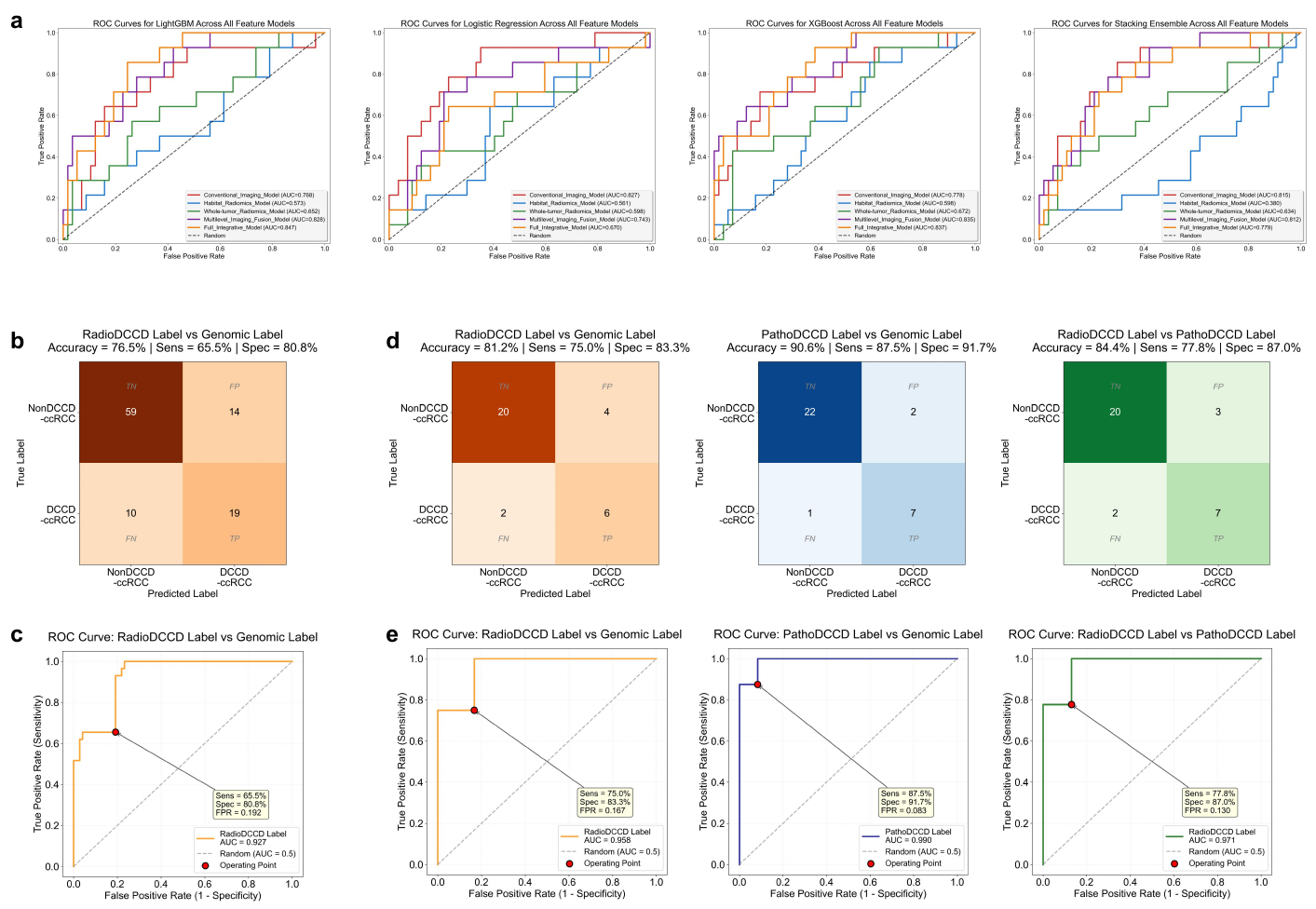}
  \caption{Development and validation of a RadiopDCCD for discriminating DCCD molecular subtypes in ccRCC. (a) The ROC curve of the RadiopDCCD classifier constructed using 20 selected radiomics features from CTA scans in Center 3, evaluated via five-fold stratified cross-validation (AUC = 0.92). (b) SHAP summary plot illustrating the importance of individual radiomic features in the model. (c–d) External validation of the model in Center 1: (c) accuracy, (d) ROC curve (AUC = 0.74), and (e) precision–recall curve. (f–g) Kaplan–Meier survival analysis in Centers 4 and 5 showing significant differences in DFS (f) and OS (g) between DCCD-rDCCD and NonDCCD-rDCCD groups. Log-rank P values are indicated for each comparison.  \\
\textbf{Abbreviations}: DCCD-ccRCC: de-clear cell differentiated clear cell renal cell carcinoma; NonDCCD-ccRCC: non-de-clear cell differentiated clear cell renal cell carcinoma; ROC: receiver operating characteristic; AUC: area under the curve; DFS: disease-free survival; OS: overall survival.
}\label{fig.05}
\end{figure}

\begin{figure}[htpb!]
  \centering
         \includegraphics[width=\linewidth]{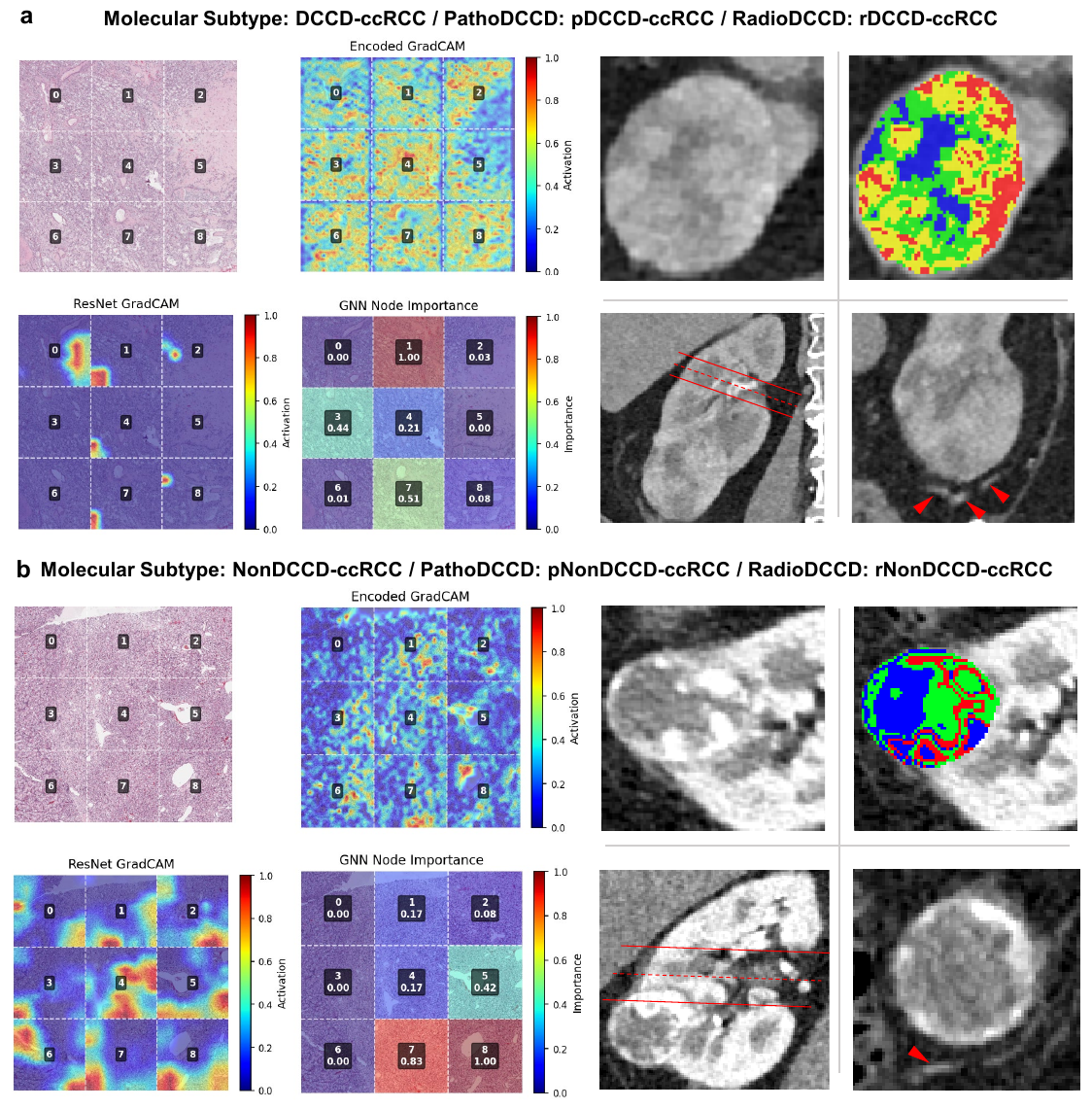}
  \caption{Multi-modal interpretability analysis of deep learning models for discriminating DCCD-ccRCC and NonDCCD-ccRCC subtypes. (a) A representative DCCD-ccRCC case. The tumor presented with mild enhancement (enhancement ratio: 1.69), a high 2D ITH score (0.828), an optimal cluster number of 4, a RENAL score of 5, and a neovascularization score of 3. Interpretability maps (top to bottom) show: Encoded Grad‑CAM highlighting spatially concentrated regions in the fused feature space; ResNet Grad‑CAM indicating focused imaging‑based attention; and GNN Node Importance revealing several high‑importance nodes (importance $\approx$1.0) in the pathological graph. The patient experienced tumor recurrence at 47 months of follow-up.
(b) A representative NonDCCD-ccRCC case. The tumor exhibited marked peripheral enhancement (enhancement ratio: 4.31), a lower 2D ITH score (0.521), an optimal cluster number of 3, a RENAL score of 8, and a neovascularization score of 1. Corresponding interpretability maps display more diffuse activations in both Grad‑CAM outputs and generally lower GNN node‑importance scores, reflecting weaker localized discriminative features. The patient remained alive without recurrence at 82 months of follow-up. Color bars denote activation intensity (0-1) for Grad-CAM and importance scores for GNN nodes.}\label{fig.06}
\end{figure}

\bibliographystyle{naturemag}
\bibliography{Reference}

\newpage
\begin{footnotesize}
\setlength{\extrarowheight}{1.1pt}
\setlength{\tabcolsep}{3.5pt}
\renewcommand{\arraystretch}{1.12}

\begin{longtable}{@{}>{\raggedright\arraybackslash}p{0.26\linewidth}
                  >{\centering\arraybackslash}p{0.11\linewidth}
                  >{\centering\arraybackslash}p{0.11\linewidth}
                  >{\centering\arraybackslash}p{0.11\linewidth}
                  >{\centering\arraybackslash}p{0.11\linewidth}
                  >{\centering\arraybackslash}p{0.11\linewidth}
                  >{\centering\arraybackslash}p{0.11\linewidth}@{}}
\caption{Demographic, Clinical, and Pathological Characteristics of Enrolled Patients at Baseline}\label{Table:01}\\
\toprule
 & \shortstack{Center 1\\(n = 478)}
 & \shortstack{Center 2\\(n = 121)}
 & \shortstack{Center 3\\(n = 353)}
 & \shortstack{Center 4\\(n = 346)}
 & \shortstack{Center 5\\(n = 329)}
 & \shortstack{Gene cohort\\(n = 32)}\\
\midrule
\endfirsthead

\toprule
 & \shortstack{Center 1\\(n = 478)}
 & \shortstack{Center 2\\(n = 121)}
 & \shortstack{Center 3\\(n = 353)}
 & \shortstack{Center 4\\(n = 346)}
 & \shortstack{Center 5\\(n = 329)}
 & \shortstack{Gene cohort\\(n = 32)}\\
\midrule
\endhead

\midrule
\multicolumn{7}{r}{\textit{Continued on next page}}\\
\midrule
\endfoot

\bottomrule
\endlastfoot

%-------------------- Age --------------------
\multicolumn{7}{@{}l}{\textbf{Age (years)}}\\
\hspace{1em}Median & 62 & 61 & 58 & 59 & 60 & 57\\
\hspace{1em}Range  & 26--86 & 30--90 & 19--88 & 28--83 & 31--85 & 30--73\\[2pt]

%-------------------- Gender --------------------
\multicolumn{7}{@{}l}{\textbf{Gender}}\\
\hspace{1em}Male   & 309 (64.6) & 93 (76.9) & 265 (75.1) & 245 (70.8) & 240 (72.9) & 21 (65.6)\\
\hspace{1em}Female & 169 (35.4) & 28 (23.1) & 88 (24.9)  & 101 (29.2) & 89 (27.1)  & 11 (34.4)\\[2pt]

%-------------------- BMI --------------------
\multicolumn{7}{@{}l}{\textbf{BMI}}\\
\hspace{1em}$\leq$ 25 kg/m$^2$ & --- & 12 (26.4) & 227 (64.3) & 210 (60.7) & 155 (47.1) & 24 (75.0)\\
\hspace{1em}$>$ 25 kg/m$^2$    & --- & 34 (73.6) & 126 (35.7) & 136 (39.3) & 174 (52.9) & 8 (25.0)\\[2pt]

%-------------------- Clinical Presentation --------------------
\multicolumn{7}{@{}l}{\textbf{Clinical presentation}}\\
\hspace{1em}Gross hematuria & --- & --- & 42 (11.9) & 55 (15.9) & 60 (18.2) & 4 (12.5)\\
\hspace{1em}Flank/back pain & --- & --- & 49 (13.9) & 40 (11.6) & 65 (19.8) & 2 (6.3)\\
\hspace{1em}Abdominal mass  & --- & --- & 8 (2.3)   & 5 (1.4)   & 12 (3.6)  & 1 (3.1)\\[2pt]

%-------------------- Medical History --------------------
\multicolumn{7}{@{}l}{\textbf{Medical history}}\\
\hspace{1em}Smoking           & --- & 46 (38.3) & 71 (20.1) & 65 (18.8) & 95 (28.9) & 4 (12.5)\\
\hspace{1em}Alcohol           & --- & 49 (40.8) & 42 (12.0) & 38 (11.0) & 58 (17.6) & 5 (15.6)\\
\hspace{1em}Hypertension      & --- & ---       & 152 (43.1) & 125 (36.1) & 170 (51.7) & 9 (28.1)\\
\hspace{1em}Diabetes mellitus & --- & ---       & 60 (17.0)  & 50 (14.5)  & 75 (22.8)  & 2 (6.3)\\
\hspace{1em}Immunotherapy history & --- & 29 (23.9) & 115 (32.6) & 90 (26.0) & 125 (38.0) & 10 (31.3)\\[2pt]

%-------------------- Pathological Characteristics --------------------
\multicolumn{7}{@{}l}{\textbf{Pathological characteristics}}\\
\hspace{1em}Tumor size (cm) & --- & 5.14 $\pm$ 4.06 & 6.15 $\pm$ 1.90 & 5.80 $\pm$ 2.10 & 6.90 $\pm$ 2.95 & 5.2 $\pm$ 2.1\\
\hspace{1em}Tumor location  &     &                  &                  &                  &                  & \\
\hspace{2em}Left            & 231 (48.3) & --- & 186 (52.7) & 180 (52.0) & 160 (48.6) & 12 (37.5)\\
\hspace{2em}Right           & 247 (51.7) & --- & 167 (47.3) & 166 (48.0) & 169 (51.4) & 20 (62.5)\\
\hspace{1em}Perirenal fat invasion        & --- & --- & 19 (5.4) & 11 (3.2) & 25 (7.6) & 2 (6.3)\\
\hspace{1em}Renal sinus invasion          & --- & --- & 15 (4.3) & 8 (2.3)  & 22 (6.7) & 1 (3.1)\\
\hspace{1em}Pelvicalyceal system invasion & --- & --- & 19 (5.4) & 10 (2.9) & 27 (8.2) & 2 (6.3)\\
\hspace{1em}Venous system invasion        & --- & --- & 21 (6.0) & 13 (3.8) & 29 (8.8) & 1 (3.1)\\
\hspace{1em}Tumor necrosis                & --- & --- & 63 (17.8) & 50 (14.5) & 85 (25.8) & 5 (15.6)\\
\hspace{1em}Sarcomatoid differentiation   & --- & --- & 7 (2.0)  & 3 (0.9)  & 12 (3.6) & 1 (3.1)\\[2pt]

%-------------------- ISUP/WHO --------------------
\multicolumn{7}{@{}l}{\textbf{ISUP/WHO}}\\
\hspace{1em}1  & 11 (2.3)  & 6 (5.0)  & 39 (11.0) & 45 (13.0) & 30 (9.1)  & 1 (3.1)\\
\hspace{1em}2  & 203 (42.5) & 58 (48.0) & 201 (57.0) & 195 (56.4) & 160 (48.6) & 19 (59.4)\\
\hspace{1em}3  & 190 (39.7) & 47 (39.0) & 95 (27.0)  & 90 (26.0)  & 115 (35.0) & 8 (25.0)\\
\hspace{1em}4  & 70 (14.6)  & 10 (8.0)  & 18 (5.1)   & 16 (4.6)   & 24 (7.3)   & 4 (12.5)\\
\hspace{1em}NA & 4 (0.8)    & ---       & ---        & ---        & ---        & ---\\[2pt]

%-------------------- T stage --------------------
\multicolumn{7}{@{}l}{\textbf{T stage}}\\
\hspace{1em}1 & 245 (51.3) & 59 (48.8) & 293 (83.0) & 295 (85.3) & 250 (76.0) & 25 (78.1)\\
\hspace{1em}2 & 58 (12.1)  & 15 (12.4) & 18 (5.1)   & 22 (6.4)   & 35 (10.6)  & 3 (9.4)\\
\hspace{1em}3 & 164 (34.3) & 45 (37.2) & 38 (10.8)  & 26 (7.5)   & 40 (12.2)  & 4 (12.5)\\
\hspace{1em}4 & 11 (2.3)   & 2 (1.6)   & 4 (1.1)    & 3 (0.9)    & 4 (1.2)    & ---\\[2pt]

%-------------------- Outcomes --------------------
\multicolumn{7}{@{}l}{\textbf{Outcomes}}\\
\hspace{1em}Recurrence/Metastasis/Dead & 182 (38.1) & 33 (27.3) & 63 (17.8) & 48 (13.9) & 78 (23.7) & 6 (18.8)\\
\hspace{1em}Disease-free survival      & 276 (57.7) & 76 (62.8) & 268 (75.9) & 288 (83.2) & 238 (72.3) & 26 (81.3)\\
\hspace{1em}Uncertain                  & 20 (4.2)   & 12 (9.9)  & 22 (6.2)  & 10 (2.9)  & 13 (4.0)  & ---\\
\hspace{1em}Disease-free survival (months) & 63 (53--69) & 64 (59--66) & 78 (66--NR) & 82 (68--NR) & 70 (52--85) & NR (68--NR)\\

\end{longtable}

\vspace{-10pt}
\noindent
\textit{Note: Tumor size are presented as mean $\pm$ standard deviation; Disease-free survival is presented as median (95\%CI); other data are presented as n (\%). NA: not available. NR: not reached.}
\end{footnotesize}

\end{document}